\documentstyle[twocolumn,aps,epsfig]{revtex}

\begin{document}
\draft
\title{Bistability and hysteresis in tilted sandpiles}
\author {Anita Mehta}
\address{ The Abdus Salam ICTP\\ Strada Costiera 11 34100 Trieste,
Italy\footnote{Present address}\\and\\S N Bose National Centre for Basic Sciences\\
Block JD, Sector III Salt Lake\\ Calcutta 700 098,
India\footnote{Permanent address}
}
\author{G. C. Barker}
\address{Institute of Food Research,\\ Norwich Research Park, Colney,\\
Norwich NR4 7UA, UK}

\maketitle

\begin{abstract}
We show that tilting a model sandpile that has dynamic disorder
leads to bistability and hysteresis at the angle of repose. Also
the distribution of {\it local slopes} shows an interesting
dependence on the amount of tilt - weakly tilted sandpiles retain
the quasi-continuous distributions of the steady state, while
large tilt makes the distribution more discrete, with local slopes
clustered round particular values. These observations are used to
explain recent experimental results on avalanche shapes; we give a
theoretical framework in terms of directed percolation.
\end{abstract}

The angle of repose of a sandpile, $\theta_r$, is the maximum
inclination of the free surface of a stationary pile. It is well
known \cite{br} that sandpiles exhibit bistable behaviour at and
around this angle; this corresponds to a range of values for the
measured angle of repose which varies as a function of different
configurational histories. It is conventional to define this range
in terms of another angle, $\theta_m$, called the {\it angle of
maximal stability}; this is the minimum value of the angle of
the sandpile at which avalanching is inevitable. Clearly, $\theta_m >
\theta_r$ and the range $\theta_m$ - $\theta_r$ corresponds to a
range of angles between the free sandpile surface and the
horizontal such that {\it either} a stable stationary state {\it
or} avalanching can result depending on how the sandpile was
produced.

In this Letter, we show that such bistable behaviour is obtained
when we `tilt' a model sandpile that has dynamic disorder
\cite{ca}. Our findings on the correlation between avalanche
shapes, and the angle of tilt of the underlying sandpile surfaces,
match recent experimental results \cite{dd}; additionally, we
provide a theoretical explanation for our results in terms of
directed percolation.

Our model sandpile is a two-dimensional version of an earlier
disordered and nonabelian cellular automaton model \cite{ca}. The
`grains' are square prisms with dimensions $1 \times  1 \times
\alpha$, and these are placed on sites $i,j$ of a square lattice
with $ 1 \le i,j \le L$, where $L$ is the system size. A grain
within column $i,j$ may rest on  either its square ($1 \times 1$)
face or a rectangular ($1 \times \alpha$) face. We denote these
two states pictorially by $-$ or $|$ because they contribute
respectively $\Delta z = \alpha, 1$  to the total columnar height
$z(i,j)$.

Grains are deposited on the sandpile with a given probability of
landing in the $-$ or the $|$ orientation. The square face down
($-$) configuration of grains is considered to be more stable and
this implies that in general, and certainly well away from the
surface, grains contribute $\Delta z = \alpha$ to the column
height. However incoming grains, as well as all other grains in
the same column, can `flip' to the other orientation with
probabilities:
\begin{eqnarray}
P(- \rightarrow |) &=& exp (-d/d_-)\nonumber\\
P(| \rightarrow -) &=& exp (-d/d_|)
\end{eqnarray}
where $d_-$, $d_|$  are scale heights. This 'flip' embodies the
elementary excitation involved in the collective dynamics of
clusters since, typically, clusters reorganise by grain
reorientation. The depth dependence reflects the fact that surface
deposition is more likely to cause cluster reorganisation near the
surface, than deep inside the sandpile. After deposition and
possible reorganisation, each column has a local slope $s(i,j)$
given by:
\begin{equation}
s(i,j) = z(i,j) - \frac{1}{2}(z(i+1,j) + z(i,j+1))
\end{equation}

If $s(i,j) > s_c$, where $s_c$ is the critical slope, then the two
uppermost grains fall from column $i,j$ onto its neighbours
\cite{kadanoff} or, when $i$ or $j = L$, exit the system. This
process could lead to further instabilities and hence avalanching.

Since the critical slope is the threshold for permissible slopes,
it is a strong determinant of the allowable granular
configurations in the sandpile, and thus a determinant of the
relative populations of the $-$ and $|$ states. It is plausible
that $ s_c$ is inversely proportional to the `tilt' (the angle
made by the base of a sandpile with the horizontal). The greater
the tilt the more unstable will be the pile to deposition, with a
consequent decrease in $s_c$ (the critical height for column
height differences).

In the following, we fix $\alpha, d_-$ and $d_|$, and examine the
effect of varying $s_c$ on the mean slope  $<s>$ of the sandpile:
\begin{equation}
<s> = \frac {\Sigma_{i,j}z(i,j)}{L^2(L+1)}
\end{equation}
This slope corresponds to the macroscopic slope tan$\theta_r$
measured in experiments; its variation with $s_c$ is shown in Fig.
\ref{one} as a line of crosses. We note that tan$\theta_r$
decreases proportionately with $s_c$. Since the free surface of
the sandpile retains an approximately constant angle with the
horizontal, tilting it through larger angles (i.e. decreasing
$s_c$) would result in lower measured angles of repose $\theta_r$,
as observed.

Next, we mimic the effect of suddenly tilting a sandpile constructed
with a critical slope of $s_c$ by reducing this to some $s_c'< s_c
$. A set of stable slopes becomes unstable, avalanching occurs,
and the sandpile stabilises to a new mean slope $<s>$ - except
where the angle of tilt is so large that spontaneous flow occurs.
(Spontaneous flow, shown by the full line in Fig. \ref{one}, sets
in for critical slopes $s_c' \le 2\alpha$, since such  thresholds
make even ordered stackings of flat, i.e. `$-$', grains, unstable). Also, in
Fig. \ref{one}, we show  by triangular and circular symbols
(corresponding to triangular and uphill avalanches \cite{dd}) the
nature of avalanche `footprints' generated when a steady-state
sandpile is tilted through different angles and then perturbed by
additional deposition. For example, when a sandpile built with
$s_c = 2.05$ is tilted so that $s'_c \sim 1.75$, avalanches
generated by further deposition are, on average, triangular in
shape (Fig. \ref{two}a). Beyond this value of $s'_c$, uphill
avalanches result (Fig. \ref{two}b). Our results are in
qualitative agreement with recent experiments \cite{dd}.

Also, we observe that sandpiles constructed at a given tilt (with
$s_c=1.85$, say) have a  lower angle of repose than those
constructed with smaller tilt (corresponding to, say $s_c=2.05$)
that are subsequently tilted to the {\it same} angle; in this case
the respective values of the mean slope are $<s>=1.21$ and $1.39$.
Since the angle of repose represents the threshold for stationary
behaviour in a sandpile, this is a strong indicator of {\it
bistable} behaviour ; additionally, hysteresis is manifest in
these results since they show that the sandpile can {\it either be
stationary or avalanching}, depending on its configurational
history, for tan$\theta_r \in [1.21,1.39]$.

Hysteresis also manifests itself in the distributions of local
slopes for tilted piles (Fig. \ref{three}). Fig. \ref{three}a
shows the smooth distribution of local slopes obtained when a pile
is constructed in the steady state with  $s_c=1.85$. Fig.
\ref{three}b shows the much more discontinuous values of local
slopes obtained when after construction with $s_c=2.05$, the
sandpile is tilted to $s'_c=1.85$. Finally, Fig. \ref{three}c
shows the extremely jagged distributions of local slopes obtained
for even larger tilt { $s'_c=1.75$}. Clearly, different
preparation histories of our model sandpile in the {\it same}
macroscopic state can lead to extremely different surface
conformations, as in reality \cite{br}.

The observation that steady-state sandpiles have a smooth
distribution of local slopes, while tilted sandpiles manifest a
{\it clustering} of values of local slopes is easily explained.
The upper layers of a steady-state sandpile subject to sudden
tilt, are `avalanched' away, exposing bulk grains to form the new
surface. These tend to be in the $-$ orientation, with $\Delta z =
\alpha$; hence slope differences tend to be clustered around
multiples of $\alpha/2$. (This is in clear contrast to the
situation at the surface of {\it steady-state sandpiles}, where
$|$ and $-$ orientations are stochastically generated due to
deposition and rearrangement). The effect of tilting the sandpile
through larger angles is to bring grains to the surface  which
were ever deeper in the bulk; these tend to be more and more
ordered as a function of their depth \cite{ca}, leading to
increasingly discrete  distributions of local slopes.

Only minor changes in the continuous distribution of local slopes
occur in the limit of small tilts. An avalanche generated by
deposition in these circumstances will propagate only a small
distance up the slope (typically after traversing only a few
sites) until a local slope is encountered for which the movement
of a grain is insufficient to achieve the toppling threshold. The
resulting avalanches are `triangular' \cite{dd}; here, slope
instabilities propagate downwards most of the time. For sandpiles
tilted through large angles, on the other hand, only a few
discrete values of local slopes  are permissible
(Fig.\ref{three}c); the propagation of grains disturbs the
stability of grains uphill of the initiation site with a greater
probability, causing the propagation of `large' \cite{ca} or
`uphill' \cite{dd} avalanches.

Two possible measures of these distinct avalanche morphologies are
shown in Fig. \ref{four}, as a function of  $s'_c$. The crosses
denote the average fraction of the avalanche footprint that is
uphill from its point of initiation $i_0, j_0$, while the dots
denote the average distance of the centre of the avalanche below
the point of initiation. We see that there is a sharp transition,
as shown by both indicators, between triangular and uphill
avalanches at $s'_c \approx 1.8$. For  lower tilt angles (higher
$s'_c$), the avalanche centres are well below the point of
initiation, and the fraction of the avalanche uphill from this is
small. For higher tilt angles (lower $s'_c$), the avalanche
centres are much closer to the initiation point, thus revealing
that a substantial fraction of the avalanche  is uphill
\cite{below}. Similar data, for other values of the initial base
angle $s_c$, were used to construct the `phase diagram' in Fig.
\ref{one} distinguishing the two avalanche morphologies. It is
easy to see  why no  (discrete) uphill avalanches are observed for
sandpiles prepared with initially large tilt ($s_c \approx 1.9$ in
Fig. \ref{one}); further large tilts simply bring the sandpile to
its maximal angle of stability with respect to the horizontal,
$\theta_m$. Denoted by the full line in Fig. \ref{one}, this is
the angle at which a stationary sandpile becomes unstable;
spontaneous flow sets in, with continuum avalanches resulting from
the destabilisation of {\it all} sites.

Differences in the local slope distributions before and after
tilting, (Fig. \ref{three}) have implications for surface
roughness in sandpiles \cite{bmmb,pre}. The quasi-continuous
distribution of local slopes (Fig. \ref{three}a) of steady-state
sandpiles result in a more continuous, locally `smoother', surface
with a large number of disordered grain configurations. On the
other hand a more discontinuous, locally `rougher', surface
results when there are very few permissible local slopes (Fig.
\ref{three}c) as in the case of the nearly ordered configurations
exposed when sandpiles are strongly tilted. However, it appears
\cite{pre} that the disordered surfaces  of weakly tilted
sandpiles are {\it globally rougher}, despite being locally
smoother in the sense of having a continuous distribution of local
slopes. In contrast, the ordered surfaces of sandpiles submitted
to large tilt  are {\it globally smoother}, despite being locally
rougher, with a discrete distribution of local slopes. This
preliminary indication of {\it anomalous roughening} \cite{krug}
in sandpile automata can be explained, tentatively, in terms of
local slope variations.

When disorder is present, the sandpile surface can 'wander' over
length scales corresponding to many columns by making small
changes in local slope, thus adding  to the surface width without
compromising stability. However,  in  ordered sandpiles, the
allowable changes in local slopes are comparatively large and the
width cannot increase indefinitely without making the sandpile
unstable. Our observation of an abrupt decrease \cite{unpub} in
the surface width of the sandpile  at the dotted line separating
triangular and uphill avalanches in Fig. \ref{one}, supports this
argument.

Lastly, we provide a theoretical framework for the onset of uphill
avalanches in terms of directed percolation. Uphill avalanches
occur when the bonds that transfer slope  instabilities are
connected in an upwards direction; it is worth noting that
although the appropriate percolation threshold should be that of
{\it directed} bond percolation ($0.64$), the pre-existence of
downhill avalanches before this onset  implies that the the
backbend length \cite{stauffer} is infinite in the downward
direction. Hence \cite{barma} the   percolation threshold  is
reduced to $0.5$, the {\it undirected} bond percolation threshold
for square lattices. From Eq. 2, it can be shown that an
instability at  site $(i,j)$ is transferred uphill when the height
decrease $\delta z(i,j)$ at that site satisfies one of the two
following conditions :
\begin{eqnarray}
\delta z (i,j) > 2 (s_c - s(i + 1, j))\nonumber\\ \delta z (i,j)
>2 (s_c -s(i, j+1))
\end{eqnarray}

This can be used to write, in mean field, the compound probability
$T^{\uparrow}$ for the uphill transfer of instability
\begin{eqnarray}
T^{\uparrow}& = & p(\delta z=2)p(s_c - s < 1)\nonumber\\ & + &
p(\delta z=1+\alpha)p(s_c - s < \frac{1}{2} (1+\alpha))
\nonumber\\ & + & p(\delta z=2\alpha)p(s_c - s < \alpha)
\end{eqnarray}
where the $p$'s represent the probabilities of the events in
parentheses. These probabilities can be computed directly from our
simulations, and we find that $T^{\uparrow}$ varies
near-monotonically between   $0.45$ (for tilt angles where
triangular avalanches are predominant) and $0.55$ (for tilt angles
where uphill avalanches are the norm). We suggest that the
transition to predominantly uphill avalanches occurs when
$T^{\uparrow} = 0.5$, i.e. {\it when there is an infinite cluster
of bonds connecting unstable sites uphill from the point of
initiation}. This is in rough agreement with our simulation
results; better agreement can only be obtained by going beyond
mean field and including fluctuations.

\acknowledgements GCB acknowledges support from the Biotechnology
and Biological Sciences Research Council, UK ($218$/FO$6522$).

\begin{figure}
\caption{ A stability diagram for two-dimensional sandpiles with
$L=32$, $\alpha = 0.7$, $d_- =2$ and $d_| =20$. The measured mean
slope $<s>$ is plotted against the critical slope $s_c$, which
should be interpreted as an inverse tilt (see text). The crosses
(x) represent steady-state stability, the full line represents the
spontaneous flow threshold, and the broken line separates a region
of triangular, predominantly downhill avalanches ($\triangle$)
from that of uphill avalanches ($\bullet$) } \label{one}
\end{figure}

\begin{figure}
\caption{ Contour plots showing the frequency with which
avalanches initiated near the centre of a sandpile cover its
different regions. The piles have $L=50$, $\alpha = 0.7$, $d_-
=2$, $d_| =20$, $s_c = 2.05$ and (a) $s'_c= 1.95$ (b) $s'_c=
1.75$. The contours correspond to $20, 40, 60,$ and $80$ per cent
coverage, and the images are built from $\approx 3500$ test
events. } \label{two}
\end{figure}

\begin{figure}
\caption{ Distribution functions for local slopes in disordered
sandpiles with $L=50$, $\alpha = 0.7$, $d_- =2$, and $d_| =20$.
(a) A steady state pile with critical slope $s_c = 1.85$. (b) A
pile subjected to small tilt; $s_c = 2.05$ and $s'_c= 1.85$. (c) A
pile subjected to large tilt; $s_c = 2.05$ and $s'_c= 1.75$.}
\label{three}
\end{figure}

\begin{figure}
\caption{ The mean distance ($\bullet$) between the centre of the
avalanche and its point of initiation, and the average fraction of
avalanche sites uphill from the initiation point (x), plotted
against the degree of tilt for a  model sandpile with parameters
$L=50$, $\alpha = 0.7$, $d_- =2$, $d_| =20$ and $s_c = 2.05$. }
\label{four}
\end{figure}

\newpage
\epsfig{file=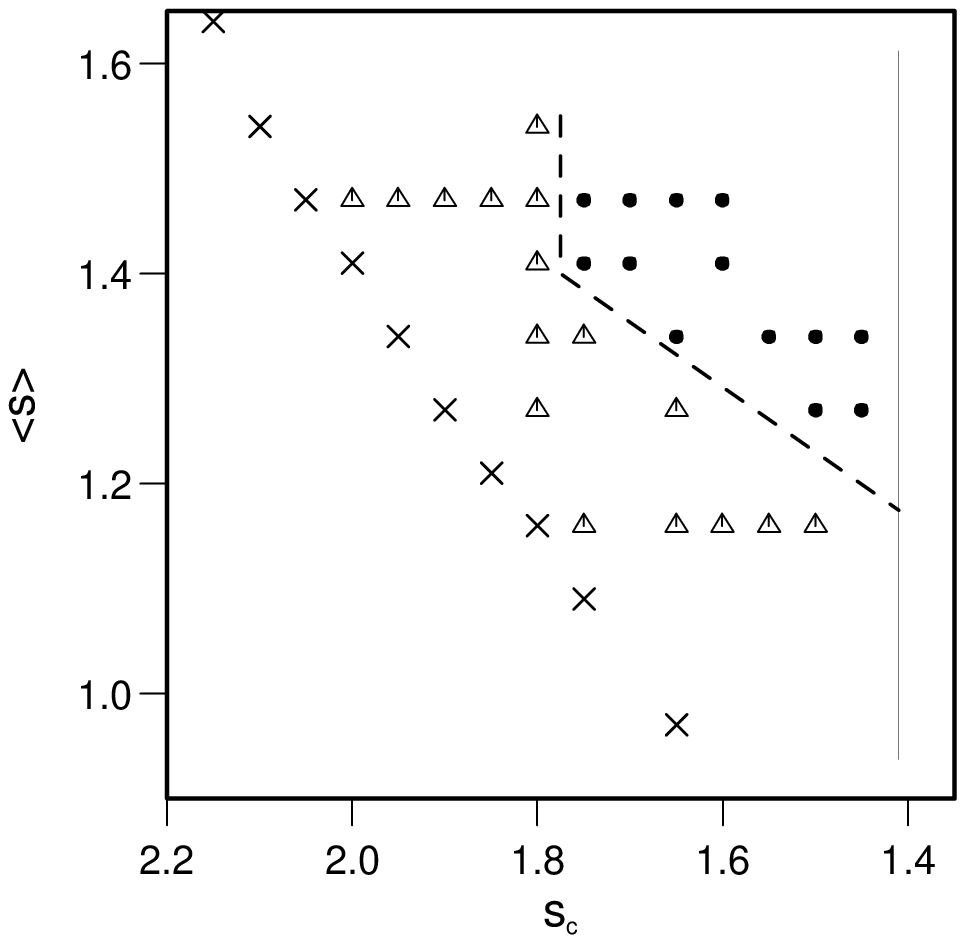,width=4in,clip=}

\newpage
\epsfig{file=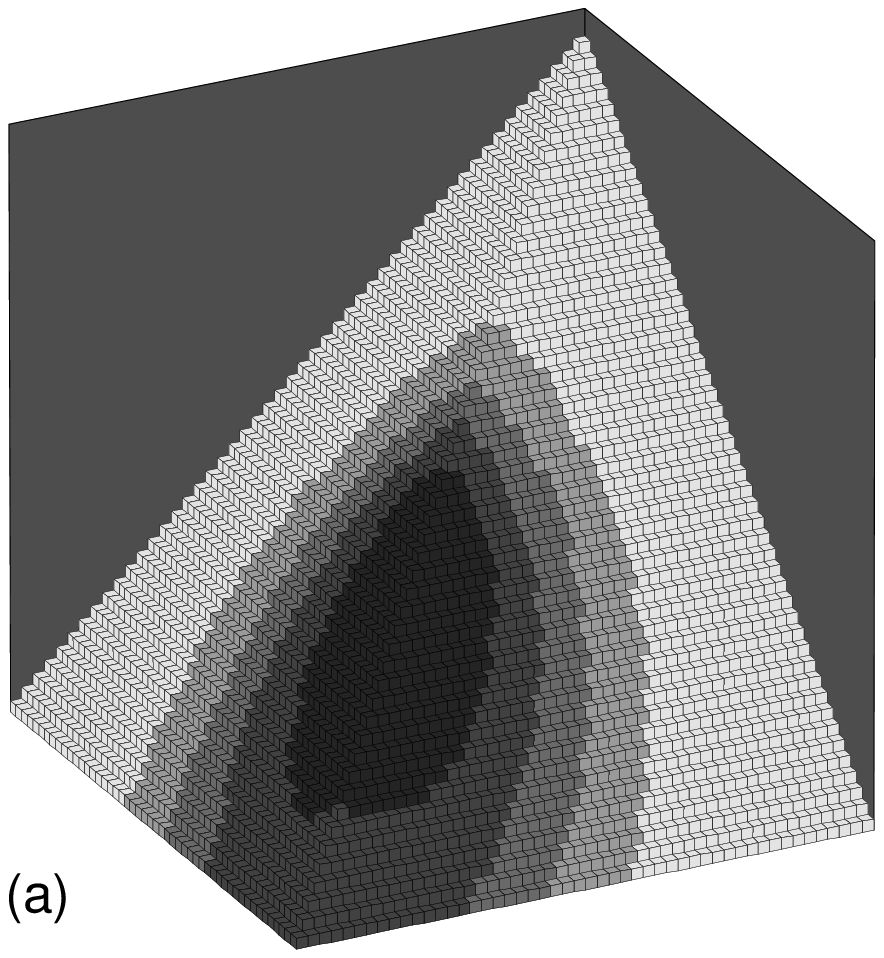,width=3in,clip=}
\epsfig{file=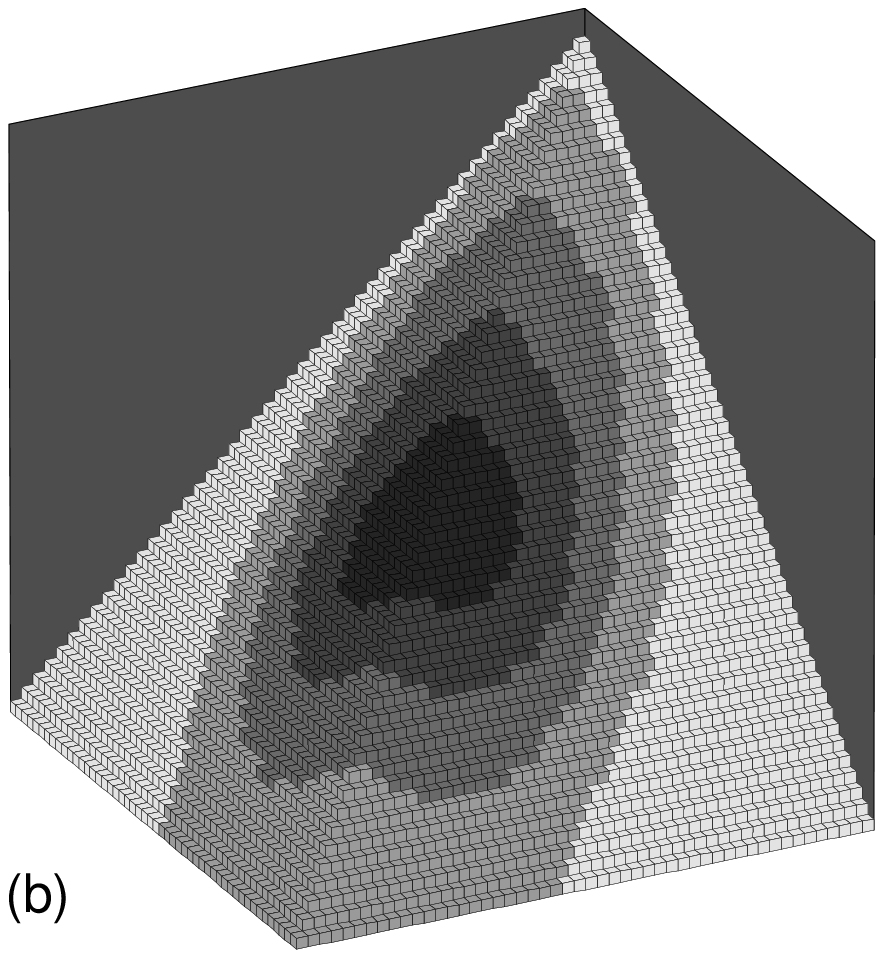,width=3in,clip=}

\newpage
\epsfig{file=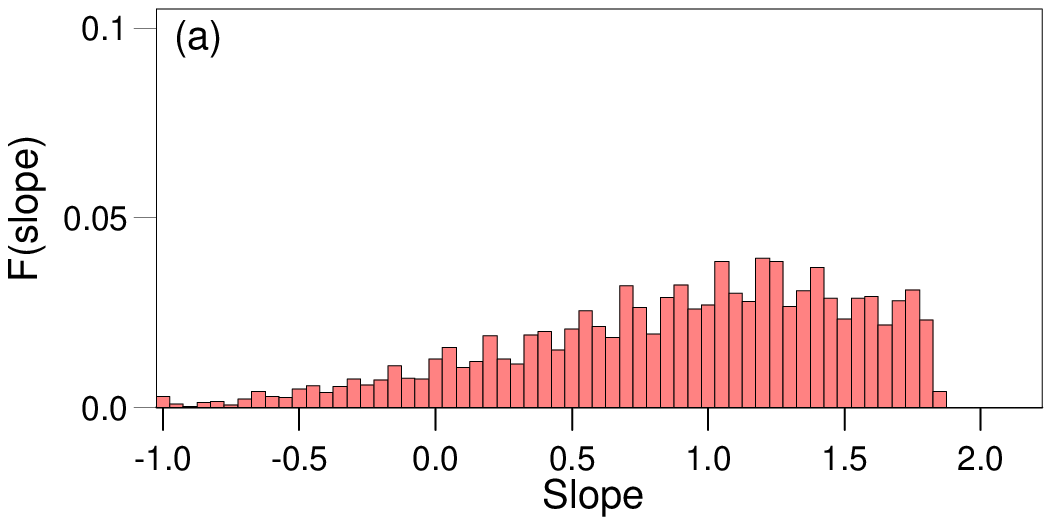,angle=270,width=1.5in,bbllx=1in,bblly=1in,bburx=3in,bbury=3in}

\epsfig{file=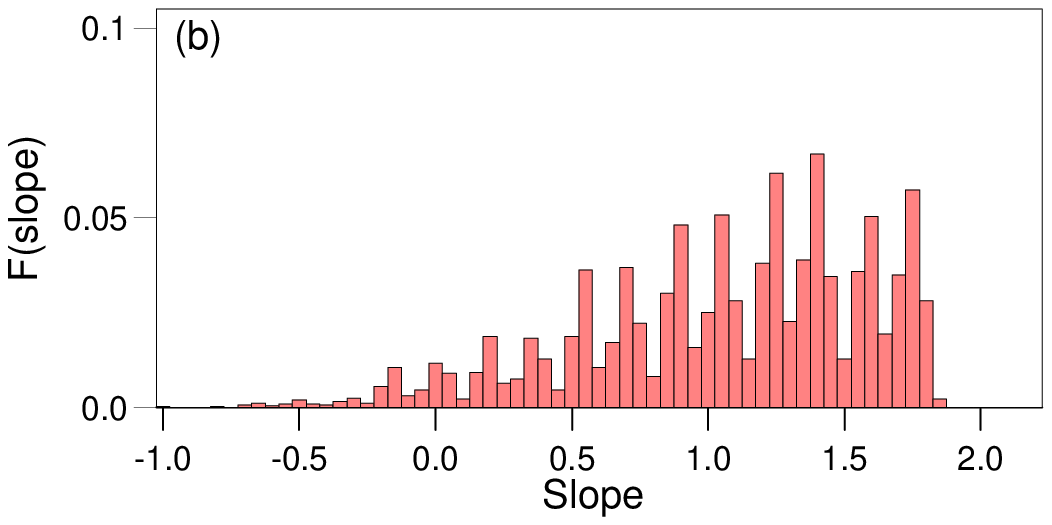,angle=270,width=1.5in,bbllx=5.25in,bblly=1in,bburx=7.25in,bbury=3in}

\epsfig{file=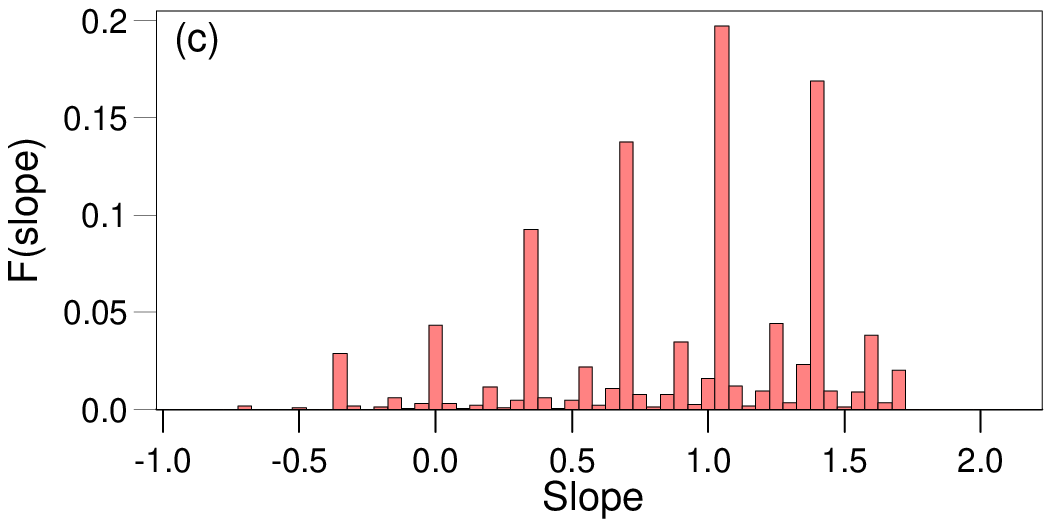,angle=270,width=1.5in,bbllx=9.5in,bblly=1in,bburx=11.5in,bbury=3in}

\newpage
\epsfig{file=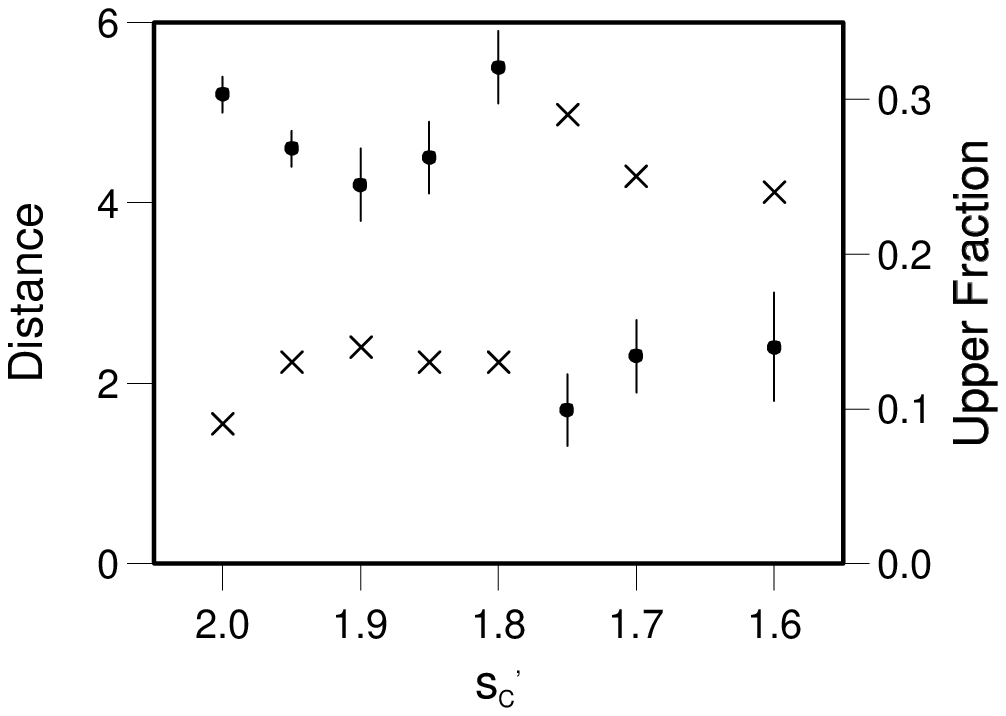,width=4in,clip=}

\end{document}